\documentstyle[12pt]{article}

\newcommand{\beq}{\begin{equation}}
\newcommand{\eeq}{\end{equation}}
\newcommand{\beqa}{\begin{eqnarray}}
\newcommand{\eeqa}{\end{eqnarray}}
\newcommand{\beqar}{\begin{eqnarray*}}
\newcommand{\eeqar}{\end{eqnarray*}}

\newcommand{\Ga}{\Gamma}

\newcommand{\inn}{\!\cdot\!}

\renewcommand{\l}{\lambda}
\renewcommand{\L}{\Lambda}

\newcommand{\sig}{\sigma}

\newcommand{\z}{\zeta}

\newcommand{\eg}{{\it e.g.,}\ }
\newcommand{\ie}{{\it i.e.,}\ }
\newcommand{\labell}[1]{\label{#1}} 
\newcommand{\reef}[1]{(\ref{#1})}
\newcommand\prt{\partial}

\newcommand\ls{\ell_s}
\newcommand\cF{{\cal F}}
\newcommand\cL{{\cal L}}
\newcommand\cG{{\cal G}}

\newcommand\bz{\bar{z}}

\newcommand\hF{\hat{F}}
\newcommand\hA{\hat{A}}

\newcommand\hD{\hat{D}}

\newcommand\hP{\hat{\Phi}}
\newcommand\hS{\hat{S}}

\newcommand\Tr{{\rm Tr}}

\begin{document}

\thispagestyle{empty}
\rightline{\small hep-th/0011147 \hfill IPM/P-2000/050}
\vspace*{1cm}

\begin{center}
{\bf \Large 
Transformation of the Dirac-Born-Infeld action \\[.25em]
under the Seiberg-Witten map  }
\vspace*{1cm}

{Mohammad R. Garousi\footnote{E-mail:
garousi@theory.ipm.ac.ir}}\\
\vspace*{0.2cm}
{\it Department of Physics, University of Birjand, Birjand, Iran}\\
\vspace*{0.1cm}
and\\
{\it Institute for Studies in Theoretical Physics and Mathematics IPM} \\
{P.O. Box 19395-5746, Tehran, Iran}\\
\vspace*{0.4cm}

\vspace{2cm}
ABSTRACT
\end{center}
We explicitly  evaluate the disk S-matrix elements of one closed 
string and an arbitrary number of open string states in the presence
of a large background B-flux. From this  calculation, we show that 
in the world-volume action of D-branes in terms of non-commutative 
fields, the closed string fields must be treated as functionals of 
the non-commutative gauge fields. We also find the generalized 
multiplication rule $*_N$ between $N$ open string fields on the world-volume 
of the D-brane. In particular, this result indicates that the 
difference between the familiar $*$  and the $*_N$ product is 
just some total derivative terms. 
We show that the $*_N$ product and the dependence of the closed string 
fields on the non-commutative gauge fields emerge also from  transforming 
the ordinary Dirac-Born-Infeld action(including the closed string fields) 
under  the Seiberg-Witten map. We then conjecture a non-commutative DBI 
action  for the transformation of the commutative  DBI action 
under the SW map.

\vfill
\setcounter{page}{0}
\setcounter{footnote}{0}
\newpage

\section{Introduction} \label{intro}

Recent years have seem dramatic progress in the understanding
of non-perturbative aspects of string theory -- see, \eg  \cite{excite}.
With these studies has come the realization that solitonic extended
objects, other than just strings, play an essential role.
An important object in these investigations has been Dirichlet
branes \cite{joep}. D-branes are non-perturbative states 
whose perturbative excitations are described by the fundamental 
open string states on their world-volumes. They also couple 
to various perturbative closed string states  including the 
Ramond-Ramond states.

The low energy effective world-volume theory of a single D-brane is a $U(1)$ gauge theory.  An interesting extension of a gauge theory is its generalization to non-commutative  theory. Recently, it has been  shown that certain non-commutative gauge  theor
ies emerge naturally  
as the D-brane world-volume theory when a constant NS-NS background B-flux is turned on \cite{acmrd,mrd,sw, RG, NS, RGS, EB}. Hence, one may study the D-brane with the B-flux to learn more about the non-commutative gauge theories. An aspect of non-commuta
tive gauge theories that the D-brane study reveals is that the multiplication rule between gauge fields in general is not just the $*$ product\cite{mine}. The new multiplication rule $*'$ between two gauge fields appears when one studies the S-matrix elem
ents of one  closed and two open string states at tree level.
This multiplication rule appears also in calculating  the S-matrix element of four open string states at one loop level\cite{hljm,dz,asdz}, and in studying in detail  anomalies in non-commutative gauge theories\cite{fans}. In \cite{hljm}, another  multipl
ication rule $*_3$ which operates between three  gauge fields was also found.

The  disk S-matrix elements of one closed and two open string states in the presence of an arbitrary  background B-flux was evaluated in \cite{mine}.
In the present paper, we would like to extend that calculation  to the case of  one closed string and an arbitrary number of open string states in the presence of a large background B-flux. From this calculation, we find  multiplication rules $*_N$ betwee
n $N$ open string fields.  An interesting aspect of the new multiplication rule is that the difference between $*_N$ and $*$ is some total derivative terms. Hence, in the actions that involve both open and closed string fields, one must apply $*_N$ as the
 multiplication rule between  open string fields. 
Whereas, in the tree level action that involves only the open string fields, like the non-commutative DBI action \cite{sw}, this $*_N$ can be replaced with the $*$ product upon ignoring some total derivative terms. Moreover, in the one loop effective acti
on of non-planer diagrams the $*_N$ dose not operate between all the fields in the action, so the total derivative terms can not be ignored. Finally, it explains the appearance of $*$ product in the one loop effective action of planer diagrams in which $*
_N$ operates between all the gauge fields in the action\cite{hljm}.

Another fact that the above S-matrix elements reveals is that the closed string fields in the non-commutative BDI action should be treated as functional of the non-commutative gauge fields. In particular, the S-matrix elements can be reproduced in field t
heory if one assumes that the closed string fields in this action are functional of the non-commutative gauge field and then Taylor expands these fields around zero gauge field.   Various terms in the Taylor expansion are proportional to    the antisymmet
ric  matrix $\theta^{ab}$ that appears in the definition of the $*$ product.  
Hence, this feature like the $*_N$ product is special to non-commutative field theories.

In \cite{mine}, it was also shown that the result of disk S-matrix elements of one closed string NSNS state and two massless open string NS states can be reproduced in field theory by the ordinary DBI action and the SW map.  One begins with expanding the 
DBI action around the background B-flux to produce an array of interactions.  Then using the SW map, one transforms the array  to non-commutative counterpart. The resulting non-commutative field theory  reproduces exactly  the above S-matrix elements. 
In particular, this calculation shows that the $*'$ product is reproduced in field theory by the SW map. In \cite{tmmw}, it was shown that the $*_3$ product is reproduced by the SW map as well.  We shall show that the various terms in the transformed DBI 
action that involve the antisymmetric matrix $\theta^{ab}$ can be  combined properly to reproduce the appropriate terms in the Taylor expansion of the closed string fields.  Hence, both  $*_N$ product and the dependence of the closed string fields on non-
commutative gauge fields seems to be reproduced in field theory by the ordinary DBI action and the SW map.

The reminder of the paper is organized  as follows:  We begin in section 2 by expanding the ordinary DBI action in Type 0 theories around the background B-flux, keeping the terms that involve one closed string tachyon and one, two or three  massless open 
string fields.
In Section 2.1 we transform these
couplings between commutative fields  
to their non-commutative counterparts using the SW map. In particular, we combine the different terms that involve the matrix $\theta^{ab}$ in a compact form. This latter form is suggestive: the closed string fields in the DBI action should be treated as 
functional of the non-commutative gauge fields. 
In Section 3, we evaluate  the disk S-matrix elements of one closed string tachyon and an arbitrary number of massless open string states in the presence of a large background B-flux. In particular, in this section we find the generalized star product $*_
N$ in momentum space, and show that the difference between $*_N$ and $*$ is some total derivative terms. The S-matrix elements are fully consistent with the coupling found in Section 2.1 as expected.
In Section 4,  using the two features of the non-commutative field theories, \ie the dependence of closed string fields on the non-commutative gauge fields and the appearance of the $*_N$ product, we conjecture a non-commutative DBI action for   transform
ation of the ordinary DBI action under the SW map.

\section{Commutative DBI action}

The world-volume theory of a single  D-brane in type 0 theory 
includes a massless
U(1) vector $A_a$ and  a set of massless scalars $X^i$, describing the transverse
oscillations of the brane\cite{leigh, irk}.
The leading order low-energy effective action for the massless
fields corresponds to a dimensional reduction of a ten dimensional
U(1) Yang Mills theory. As usual in string theory, there are
higher order $\alpha'=\ls^2$ corrections,
where $\ls$ is the string length scale. As long as derivatives
of the field strengths (and second derivatives of the scalars)
are small compared to $\ls$, then the action takes a Dirac-Born-Infeld
form \cite{bin}. 
To take into account the couplings of the massless open string states
with closed strings, the DBI
 action may be extended to include massless Neveu-Schwarz
closed string fields,\ie  the metric, dilaton and
Kalb-Ramond fields, and the closed string tachyon\cite{irkaat,mrga}. In this way  one arrives at the 
following world-volume
action:
\beqa
S&=&-T_p \int d^{p+1}\sig\,g[T(\Phi)] e^{-\phi(\Phi)}\sqrt{-{\rm det}(P[G_{ab}(\Phi)+
B_{ab}(\Phi)]+\l F_{ab}}\labell{biact}
\eeqa
where $\l=2\pi\ell_s^2$, and the closed string tachyon function is $g[T]=1+T/4+\cdots$. Here, $F_{ab}$ is the abelian field
strength of the world-volume ordinary gauge field, while
the metric and antisymmetric tensors are
the pull-backs of the bulk tensors to the D-brane world-volume, \eg
\beqa
P[G_{ab}]&=&
G_{ab}+2\l G_{i(a}\,\prt_{b)}\Phi^i+\l^2G_{ij}\prt_a\Phi^i\prt_b\Phi^j
\labell{pull}
\eeqa
where  we have used that fact that we are employing static
gauge throughout the paper, \ie $\sig^a=X^a$ for world-volume and 
$\l\Phi^i(\sig^a)$ for
transverse coordinates. In the action \reef{biact}, we include in fact derivatives of the closed string fields by writing them as functional of the transverse scalar fields\cite{hull,gm}, \eg
\beqa
T(\Phi)&=&T^0+\l\Phi^i\prt_i T^0+\frac{\l^2}{2}\Phi^i\Phi^j\prt_i\prt_j T^0+\frac{\l^3}{3!}\Phi^i\Phi^j\Phi^k\prt_i\prt_j\prt_k T^0+\cdots
\labell{taylor}\eeqa 

To study the transformation of \reef{biact} under the SW map, we begin with   
expanding \reef{biact}  
for fluctuations around the background
$G_{\mu\nu}=\eta_{\mu\nu}$, $B_{\mu\nu}=\cF^{ab}\eta_{a\mu}\eta_{b\nu}$,
$\Phi=0$ and $T=0$ to extract the couplings  expected from the DBI action.
To simplify our illustration, we ignore all the closed string fields in \reef{biact} but the tachyon, that is,
\beqa
S&=&-T_p\int d^{p+1}\sigma\,g[T(\Phi)]\sqrt{-\det(\eta_{ab}+\cF_{ab}+\l F_{ab})}\labell{biact1}\eeqa

Now  it is straightforward,
to expand eq.~\reef{biact1} using
\beqar
\sqrt{{\rm det}(M_0+M)}&\!\!\!=\!\!\!&\sqrt{{\rm det}(M_0)}\left(\frac{}{}
1+{1\over2}\Tr(M_0^{-1}M)
-{1\over4}\Tr(M_0^{-1}MM_0^{-1}M)+{1\over8}(\Tr(M_0^{-1}M))^2\right.
\nonumber\\
&&+{1\over6}\Tr(M_0^{-1}MM_0^{-1}MM_0^{-1}M)-{1\over8}\Tr(M_0^{-1}M)\Tr(M_0^{-1}MM_0^{-1}M)\nonumber\\
&&\left.+{1\over48}(\Tr(M_0^{-1}M))^3+\cdots\right)
\eeqar
to produce a vast array of interactions. We are  interested in
the coupling  linear in the tachyon field and linear, quadratic or triadic in massless open 
string fluctuations. 
The 
appropriate Lagrangian is: 
\beqa
\cL(T)&=&-\frac{T_p c}{4}T(\Phi)\nonumber\\
\cL(T,A)&=&-\frac{T_p \l c}{4}T(\Phi)\left(\frac{1}{2}\Tr(VF)
\right)\labell{int3}\\
\cL(T,2A)&=&-\frac{T_p \l^2c}{4}T(\Phi)\left(-\frac{1}{4}\Tr(VFVF)+\frac{1}{8}(\Tr(VF))^2)\right)
\nonumber
\\
\cL(T,3A)&=&-\frac{T_p\l^3 c}{4}T(\Phi)\left(\frac{1}{6}\Tr(VFVFVF)-\frac{1}{8}\Tr(VF)\Tr(VFVF)+\frac{1}{48}(\Tr(VF))^3\right)\nonumber
\eeqa
In the above 
Lagrangian, 
\beqa
c\,\,\equiv\,\,\sqrt{-{\rm det}(\eta_{ab}+\cF_{ab})}&\,\,,\,\,&
V^{ab}\,\,\equiv\,\, \left( (\eta+\cF)^{-1}\right)^{ab} \,\,, 
\labell{vmatrix}
\eeqa
Using the Taylor expansion \reef{taylor}, one includes the scalar fields in the above couplings. Now we transform these couplings to their non-commutative counterparts under the SW map.

\subsection{Change of variables}

In \cite{sw}  differential equation for transforming ordinary gauge field to its
non-commutative counterpart
was found  to be
\beqa
\delta \hA_a(\theta)\!\!\!&=\!\!\!&\frac{1}{4}\delta\theta^{cd}\left(\hA_c*\hF_{ad}+\hF_{ad}*A_c-\hA_c*\prt_d\hA_a
-\prt_d\hA_a*A_c\right)\labell{delf}\\
\delta\hF_{ab}(\theta)\!\!\!&=\!\!\!&\frac{1}{4}\delta\theta^{cd}\left(
2\hF_{ac}*\hF_{bd}+2\hF_{bd}*\hF_{ac}
-\hA_c*(\hD_d\hF_{ab}+\prt_d\hF_{ab})
-(\hD_d\hF_{ab}+\prt_d\hF_{ab})*\hA_c\right)
\nonumber
\eeqa
where the gauge field strength and $*$ product were defined to be
\beqa
\hF_{ab}&=&\prt_a\hA_b-\prt_b\hA_a-i\hA_a*\hA_b+i\hA_b*\hA_a\nonumber\\
&=&\prt_a\hA_b-\prt_b\hA_a-i[\hA_a,\hA_b]_M\nonumber\\
f(x)*g(x)&=&e^{i\prt_1\L\prt_2}
f(x_1)g(x_2)|_{x_i=x}\,\, ,\nonumber
\eeqa
where
\beqa
\prt_1\L\prt_2&=&\frac{1}{2}\theta^{ab}\frac{\prt}{\prt x_1^a}\frac{\prt}{\prt x_2^b}
\eeqa
These differential equations were solved in \cite{mine} to order $O(\hA^2)$ by integration along a special path in the space of the matrix $\theta^{ab}$. It gives  a relation
between ordinary fields appearing in \reef{int3} and non-commutative fields corresponding
to open string vertex operators. The result for abelian case that we are interested in
is 
\beqa
A_a&=&\hA_a+\frac{1}{2}\theta^{bc}\hA_b*'\left(\hF_{ca}+\prt_c\hA_a\right)+O(\hA^3)
\nonumber\\
F_{ab}&=&\hF_{ab}-\theta^{cd}\left(\hF_{ac}*'\hF_{bd}-\hA_c*'\prt_d\hF_{ab}\right)
+O(\hA^3)
\labell{fhf}
\eeqa
where the multiplication rule between two fields  operates as
\beqa
f(x)*'g(x)&=&\frac{\sin(\prt_1\L\prt_2)}
{\prt_1\L\prt_2}f(x_1)g(x_2)|_{x_i=x}\,\, .\nonumber
\eeqa

These  solutions were improved in \cite{tmmw} to include the terms of order $O(\hA^3)$, that is,
\beqa
A_a&=&\hA_a+\frac{1}{2}\theta^{bc}\hA_b*'(\prt_c\hA_a+\hF_{ca})\nonumber\\
&&+\frac{1}{2}\theta^{bc}\theta^{de}\left(-\hA_b\prt_d\hA_a(\prt_c\hA_e+\hF_{ce})+\prt_b\prt_d\hA_a\hA_c\hA_e+2\prt_d\hA_b\prt_a\hA_c\hA_e\right)*_3+O(\hA^4)
\nonumber\\
F_{ab}&=&\hF_{ab}-\theta^{cd}\left(\hF_{ac}*'\hF_{bd}-\hA_c*'\prt_d\hF_{ab}\right)\labell{trans3}\\
&&+\frac{1}{2}\theta^{cd}\theta^{ef}\left(\prt_c\prt_e(\hF_{ab}\hA_d\hA_f)-\prt_e(\hF_{cd}\hF_{ab}\hA_f)+2\prt_e(\hF_{ac}\hF_{bd}\hA_f)\right.\nonumber\\
&&\left.-(\hF_{ac}\hF_{bd}\hF_{ef})+\frac{1}{4}(\hF_{ab}\hF_{cd}\hF_{ef})+\frac{1}{2}(\hF_{ab}\hF_{de}\hF_{cf})-2(\hF_{ce}\hF_{af}\hF_{bd})\right)*_3+O(\hA^4)\nonumber
\eeqa
where the multiplication  rules between three fields is 
\beqa
\left(f(x)g(x)h(x)\right)*_3&=&\left(\frac{\sin(\prt_2\L\prt_3)}{(\prt_1+\prt_2)\L\prt_3}\frac{\sin(\prt_1\L(\prt_2+\prt_3)}{\prt_1\L(\prt_2+\prt_3)}+1\leftrightarrow 2\right)f(x_1)g(x_2)h(x_3)|_{x_i=x}\nonumber
\eeqa
The eq.~\reef{trans3} may be used to find the transformation of the scalar fields in adjoint representation of $U(1)$, for example, 
\beqa
\Phi_i&=&\hP_i+\theta^{bc}\hA_b*'D_c \hP_i+\frac{1}{2}\theta^{bc}\theta^{de}\left(
-\hA_b\prt_d\hP_i(\prt_c\hA_e+\hF_{ce})+\prt_b\prt_d\hP_i\hA_c\hA_e\right)*_3+\cdots
\labell{la2}
\eeqa
where dots represent terms which involve  more than three open string fields. 
They produce couplings between  more than three fields upon replacing 
them into \reef{int3} 
in which we are not interested in this section. 
In above equation $D_c\hP_i=\prt_c\hP_i-i[\hA_c,\hP_i]_M$.

The differential equation \reef{delf}  expresses infinitesimal variation 
 of linear field 
, \eg $\delta \hA_a$, in terms of infinitesimal variation of the 
non-commutative parameters, \ie $\delta\theta^{cd}$.
Upon integration \reef{trans3}, this transforms the linear commutative gauge 
field in terms of
nonlinear combination of non-commutative fields. 
However, we are interested in transforming
non-linear  combinations of commutative fields appearing in \reef{int3} 
in terms of non-commutative fields.
Such a transformation, in principle, might  be found from a  differential 
equation alike \reef{delf} that
 expresses infinitesimal  variation of the non-linear  fields 
in terms of infinitesimal  variation of
non-commutative parameter. Upon integration, that would produce the desired
 transformation.
In that way, one would find that not only the fields transform 
as in \reef{trans3}
but the multiplication rule between  fields also  undergo
appropriate transformation.  
We are not going to find such a differential equations here. Instead, we simply
note that  the transformation for multiplication rule between two and three open
string fields can be read from the right hand side of eq.~\reef{trans3} to be
\beqa
fg|_{\theta=0}&\longrightarrow&f*'g|_{\theta\ne 0}\nonumber\\
fgh|_{\theta=0}&\longrightarrow& (fgh)*_3|_{\theta\ne 0}
\labell{fg}
\eeqa
for  $f,g$ and $h$ being any arbitrary open string fields. 

Now with the help of equation \reef{trans3}, \reef{la2}  and \reef{fg}, one can transform 
the commutative Lagrangian \reef{int3} to non-commutative 
counterpart.
In doing so, one should first ignore the multiplication 
rules in \reef{trans3} and \reef{la2} \ie $*'$ and $*_3$. 
Then using the resulting transformations, one  maps the commutative 
couplings in \reef{int3} to their non-commutative counterparts. 
Finally,  one should replace the generalized star product  between 
the non-commutative fields $\hP^i$, $\hA_a$ or $\hF_{ab}$
\footnote{Note that $*'$ and $*_3$ products are 
invariant under all permutations of the non-commutative fields\cite{tmmw}.}.
Let us work more details on the coupling between the tachyon, 
the transverse scalar and the gauge field. One starts with the following 
couplings
\[
-\frac{T_p c}{4}\left(\l \prt_iT^0\Phi^i+\l^2\prt_iT^0\Phi^i V_A^{ab}\prt_b A_a\right)
\]
which is the first Taylor expansion of $\cL(T)+\cL(T,A)$. Now one  
transforms the commutative fields in the above couplings to their 
non-commutative fields using the transformations \reef{trans3} 
and \reef{la2}. Since we are interested in the terms with two open 
string fields, one should replace the open string fields in the second 
term above with their non-commutative fields, \ie first 
terms of \reef{trans3} and \reef{la2}. Whereas, in the first term 
the scalar field must be replaced with the second term of \reef{la2}. 
In this way one finds
\footnote{Note that our conventions set $\theta^{ab}=\l (V_A)^{ab}$ 
where $V_A$ is the antisymmetric part of the V-matrix \reef{vmatrix}.},
\[
-\frac{T_p \l^2 c}{4}\left(\prt_i T^0V_A^{ab}\hA_a\prt_b\hP^i+\prt_i T^0\hP^iV_A^{ab}\prt_b\hA_a\right)
\]
Finally, one should insert the multiplication rule 
between the two open string fields, that is 
\beqa
{\hat{\cL}}(T^0,\hP,\hA)&=&-\frac{T_p\l^2 c}{4}\prt_i\prt_aT^0(\hP^i*'\hA^a)\labell{hatl1}\\
{\hat{\cL}}(T^0,2\hA)&=&-\frac{T_p\l^2 c}{4}\left(\frac{1}{2}\prt_a\prt_bT^0(\hA^a*'\hA^b)-\frac{1}{4}T^0(V_S^{ab}\hF_{bc}V_S^{cd}*'\hF_{da})\right)\nonumber\\
{\hat{\cL}}(T^0,2\hP,\hA)&=&-\frac{T_p\l^3 c}{8}\prt_a\prt_i\prt_jT^0(\hP^i\hP^j \hA^a)*_3\nonumber\\
{\hat{\cL}}(T^0,\hP,2\hA)&=&-\frac{T_p\l^3 c}{4}\left(\frac{1}{2}\prt_a\prt_b\prt_i T^0(\hP^i \hA^a\hA^b)*_3-\frac{1}{4}\prt_i T^0(\hP^iV_S^{ab}\hF_{bc}V_S^{cd}\hF_{da})*_3\right)\nonumber\\
{\hat{\cL}}(T^0,3\hA)&=&-\frac{T_p\l^3 c}{4}\left(\frac{1}{3!}\prt_a\prt_b\prt_cT^0(\hA^a\hA^b\hA^c)*_3-\frac{1}{4}\prt_e T^0(\hA^eV_S^{ab}\hF_{bc}V_S^{cd}\hF_{da})*_3\right)\nonumber
\eeqa
where $\hA^a=V_A^{ab}\hA_b$, and $V_A(V_S)$ is the 
antisymmetric(symmetric) part of the $V$-matrix \reef{vmatrix}.
We have also listed in the above equation the results for some other 
couplings and  ignored some total derivative terms.
In the above equations, the fields inside the argument of 
${\hat{\cL}}(\cdots)$ refers to the number of fields in the first term of 
each equation.

Apart from the multiplication rules, $V_A^{ab}$ appears only in the first 
terms in \reef{hatl1}. These terms looks like the Taylor 
expansion \reef{taylor}, but for both $\hP^i$ and $\hA^a$. The appearance 
of $V_S^{ab}$ in the second term is consistent with the fact that the 
metric for the open string fields is $V_S^{ab}$\cite{sw}. Now we turn to 
the string theory side to illustrate that the above terms are produced 
by the string theory S-matrix elements, and then generalize above terms  
to $N$ open string fields.

\section{Scattering Calculations}

In this Section we evaluate  the  coupling of one closed 
string tachyon with  an arbitrary number of massless open string states 
on the world-volume of D-brane with  large background B-flux.   
The corresponding scattering amplitude is given as 
\beqa
A_{12\cdots N}&=&\frac{\l^N T_p c}{2(2\pi)^{N-1}}\int dx_1\cdots dx_Nd^2z\,\langle 0|T\left(
V_1^{NS}\cdots V_N^{NS}\,V^T\right)|0 \rangle \,\,\,,
\labell{Ansns}
\eeqa
we have normalized the amplitude such that it reproduces the 
numerical factors in \reef{hatl1}. The vertex operators above are
\beqa
V_\ell^{NS}(k_\ell,\z_\ell,x_\ell)&=&(\z_{\ell}\inn\cG)_{\mu}  :V_0^{\mu}(2k_\ell\inn V^T,x_\ell):
\qquad\ell=1,2,\cdots N\nonumber\\
V^T(p,z,\bar{z})
&=&:V_{-1}(p,z):\ :V_{-1}
(p\inn D,\bz):\labell{vertechs}\nonumber\,\,,
\eeqa
where the momenta and polarizations 
satisfy $p^2=1$, $k_{\ell}\inn V_S\inn k_{\ell}=0$ 
and $k_{\ell}\inn V_S\inn \z_{\ell}=0$. The $V_0^{\mu}$ and 
$V_{-1}$ are given as
\beqa
V^{\mu}_0(p,z)&=&(\prt X^{\mu}+ip\cdot \psi\psi^{\mu})e^{ip\cdot X}
\nonumber\\
V_{-1}(p,z)&=&e^{-\sigma}e^{ip\cdot X}
\labell{componv}
\eeqa
The vertex operators above are chosen such that they saturate the 
background super-ghost charge on the world-sheet, \ie $Q_{\sigma}=2$. 
In the open string vertex
operators, the index $\mu$ will run over the world-volume (transverse)
directions when it represents a world-volume vector (transverse scalar)
state. Here we are using the notation of ref.~\cite{mrg}. 
In particular, we have used the doubling trick \cite{scatd,ours} to 
convert the
disk amplitude to a calculation involving only the standard holomorphic
correlators 
\beqa
\langle 0|T\left( X^{\mu}(z_1)\,X^{\nu}(z_2)\right)|0\rangle &=&-\eta^{\mu\nu}\,{\rm log}(z_1-z_2)-\frac{i\pi}{2}\cF^{\mu\nu}\Theta(x_1-x_2)
\nonumber\\
\langle 0|T\left( \psi^{\mu}(z_1)\,\psi^{\nu}(z_2)\right)|0\rangle
&=&-\frac{\eta^{\mu\nu}}{z_1-z_2}
\labell{propagator} \\
\langle 0|T\left(\sigma(z_1)\,\sigma(z_2)\right)|0\rangle &=&-{\rm log}(z_1-z_2) \, \, . 
\nonumber
\eeqa
where the second term in the right hand side of the first equation 
above is nonzero when both $z_1$ and $z_2$ are on the real 
axis(the boundary of the world-sheet), and $\Theta(x)=1(-1)$ if 
$x>0(x<0)$. For more details on our conventions, we refer the interested 
reader to the Appendix of \cite{mine}. Replacing the vertex operators 
into the scattering amplitude \reef{Ansns}, one finds
\beqa
A_{12\cdots N}&=&\frac{\l^{N}T_p c}{2(2\pi)^{N-1}}(\z_1\inn\cG)_{\alpha_1}\cdots
(\z_N\inn\cG)_{\alpha_N}\int dx_1\cdots dx_Nd^2z\labell{amp123}\\
&&\langle 0|T\left(:(\prt X^{\alpha_1}(x_1)+2ik_1\inn V^T\inn\psi(x_1)\,\psi^{\alpha_1}(x_1))e^{2ik_1\inn V^T\inn X(x_1)}:\cdots\right.\nonumber\\
&&:(\prt X^{\alpha_N}(x_N)+2ik_N\inn V^T\inn\psi(x_N)\,\psi^{\alpha_N}(x_N))e^{2ik_N\inn V^T\inn X(x_N)}:\nonumber\\
&&\left.:e^{-\phi(z)}e^{ip\inn X(z)}::e^{-\phi(\bz)}e^{ip\inn D\inn X(\bz)}:\right)|0\rangle
\nonumber
\eeqa
Using the Wick theorem, one can evaluate the correlation function 
above with the help of the two-point functions  \reef{propagator}. 
When $\alpha_n$ takes value in the transverse space, in order to produce 
the Taylor expansion of the closed string field in terms of non-commutative 
scalar fields, one must chose only the $\prt X^{\alpha_n}$ of the 
corresponding open string vertex operator, and it should contracts only 
with the closed string vertex operator. That is,
\[
\frac{ip^{\alpha_n}}{(x_n-z)}+\frac{i(p\inn D)^{\alpha_n}}{(x_n-\bz)}=\frac{i(z-\bz)p^{\alpha_n}}{(x_n-z)(x_n-\bz)}
\]
where we have used the identity $(p\inn D)^{\alpha_n}+p^{\alpha_n}=0$. 

When $\alpha_m$ takes value in the world-volume space, in general, the 
amplitude \reef{amp123} contains many poles and contact terms. However, 
for the special case that the V-matrix is antisymmetric, \ie large 
background B-flux, the amplitude simplifies considerablely. First 
we consider the correlation between the world-sheet fermions.  
The contractions between $\psi$'s  yield different  terms which are 
proportional to one of the following terms:
\[
\z_i\inn \cG\inn V\inn \z_i,\,\,\,\,\,\,\,\,\,\,\z_i\inn\cG\inn V\inn k_j,\,\,\,\,\,\,\,\,\,\, k_i\inn \cG\inn V\inn k_j\,\,.
\]
However,  using the identity $\cG\inn V=V_S$ 
(see the Appendix in \cite{mine}), these terms do not appear for 
the case that we are interested in, \eg $V_A$. Next we consider the 
$\prt X^{\alpha_m}$ part of the open string vertex operators. 
Contractions of these parts among the open string vertex operators result 
terms of the above form which have zero effect. Whereas, their contractions  
with the closed string vertex operator yield the following terms: 
\[
\frac{ip^{\alpha_m}}{(x_m-z)}+\frac{i(p\inn D)^{\alpha_m}}{(x_m-\bz)}=\frac{i(z-\bz)p^{\alpha_m}}{(x_m-z)(x_m-\bz)}
\]
where we have used 
the conservation of momentum in the world-volume directions  
\[
\sum_{\ell=1}^{N}2k_{\ell}\inn V^T+p+p\inn D=0
\]
and used the fact that  the first term has zero result when 
V-matrix is antisymmetric. Finally, it is straightforward exercise to 
evaluate the correlations  between the exponential operators 
in \reef{amp123} using the two-point functions in \reef{propagator}. 
The final result is,  
\beqa
A_{12\cdots N}&=&\frac{\l^{N}T_p c}{2(2\pi)^{N-1}}\prod_{\ell=1}^{N}(\z_{\ell}\inn\cG\inn p)
\int dx_1\cdots dx_N d^2z
\left(\prod_{i<j}^{}e^{-i\pi l_{ij}\Theta(x_i-x_j)}\right)\nonumber\\
&&\times i^N(z-\bz)^{N-2}
\prod_{i=1}^{N}(x_i-z)^{-1+\sum_{j=1}^{j=N}l_{ij}}(x_i-\bz)^{-1-\sum_{j=1}^{j=N}l_{ij}}
\labell{amptwo}
\eeqa
where $l_{ij}=2k_i\inn V_A\inn k_j$.
A nontrivial check of the result in eq.~\reef{amptwo}
is that the integrals is $SL(2,R)$ invariant.
To remove the associated divergence and properly evaluate the amplitude,
we fix: $z=i$, $\bz=-i$, $x_1=R\rightarrow \infty$.
With this choice, one finds
\beqa
A_{12\cdots N}&=&\frac{\l^{N}T_p c}{4\pi^{N-1}}\prod_{\ell=1}^{N}(\z_{\ell}\inn\cG\inn p)
\int_{-\infty}^{+\infty} dx_2\int_{x_2}^{+\infty}dx_3\cdots\int_{x_{N-1}}^{+\infty} dx_N \nonumber\\
&&\left(\prod_{i<j}^{}e^{-i\pi l_{ij}\Theta(x_i-x_j)}\right)
\prod_{i=2}^{N}(x_i-i)^{-1+\sum_{j=1}^{j=N}l_{ij}}(x_i+i)^{-1-\sum_{j=1}^{j=N}l_{ij}}
\labell{ampfour}
\eeqa
It is not difficult to check that the amplitude above has no pole. 
Hence, this amplitude defines the  {\it coupling} between one closed 
string tachyon and $N$ massless open string fields when the background 
B-flux is large. The integral in the amplitude is then related to the 
multiplication rule between the $N$ external open string fields.

The amplitude \reef{ampfour}  is  one ordering of the external open 
string states, \ie $(23\cdots N1)$. Adding all non-cyclic permutation 
of the external open string states, one finds the full coupling between 
the tachyon and $N$ massless open string fields in the momentum space, 
\beqa
A&=&\frac{\l^{N}T_p c}{4n!m!}\left(\prod_{\ell=1}^{N}(\z_{\ell}\inn\cG\inn p)\right)*_N
\labell{ampthree}
\eeqa
where we have also divided the amplitude by $n!m!$, where $n$ and $m$ 
are the number of open string scalar and gauge fields, respectively. 
The multiplication rule between the $N$ open string fields in the momentum 
space is then 
\beqa
\left(f_1\cdots f_N\right)*_N&=&\frac{1}{\pi^{N-1}}
\sum_{}\int_{-\infty}^{+\infty} dx_2\int_{x_2}^{+\infty}dx_3\cdots\int_{x_{N-1}}^{+\infty} dx_N \nonumber\\
&&\left(\prod_{i<j}^{}e^{-i\pi l_{ij}\Theta(x_i-x_j)}\right)
\prod_{i=2}^{N}(x_i-i)^{-1+\sum_{j=1}^{j=N}l_{ij}}(x_i+i)^{-1-\sum_{j=1}^{j=N}l_{ij}}
\labell{starn}
\eeqa
where the summation is over all permutations of $2,3,\cdots, N$, 
\eg non-cyclic permutation of $1,2,\cdots, N$. An alternative formula 
for the $*_N$ was also found in \cite{hljm}.
When the background B-flux vanishes, the above multiplication rule reduces 
to ordinary product between $N$ fields, that is
\beqa
Lim_{{}_{l_{ij}\longrightarrow 0}}\left(f_1\cdots f_N\right)*_N&=&1
\eeqa
It should not be hard to perform the integrals in \reef{starn} for 
any $N$, however, we perform the calculations for  $N=2,3$. The results are
\beqa
\left(f_1f_2\right)*_2&\!\!\!=\!\!\!&\frac{\sin(\pi l_{12})}{\pi l_{12}}\,=\,f_1*'f_2\labell{star2}\\
\left(f_1f_2f_3\right)*_3&\!\!\!=\!\!\!&\frac{-1}{4\pi^2}\sum\left(\frac{e^{i\pi(l_{21}+l_{31}+l_{23})}}{(l_{12}+l_{13})(l_{12}+l_{32})}+\frac{e^{i\pi(l_{12}+l_{13}+l_{23})}}{(l_{31}+l_{32})(l_{31}+l_{21})}+\frac{e^{i\pi(l_{12}+l_{32}+l_{31})}}{(l_{31}+l_
{32})(l_{12}+l_{32})}\right)\nonumber
\eeqa
for $N=3$, the double integral  is evaluated in the Appendix. 
This momentum representation of the $*_3$ is exactly the one 
found in \cite{hljm} in studying the one loop effective action of 
non-planer diagrams.

When the closed string tachyon has no momentum in the world-volume 
directions, \ie $p^a=0$, one expects that the $*_N$ product would  
reduce to the familiar $*$ product. This can be seen from \reef{starn} 
by noting that when $p^a$ is zero $\sum_{j=1}^{j=N}l_{ij}=0$ for any $i$. 
This follows from the conservation of momentum in the world-volume 
directions, \ie $\sum_{\ell}k_{\ell}^a+p^a=0$. Therefore,
\beqa
(f_1\cdots f_N)*_N&=&\left(\prod_{i<j}e^{-i\pi l_{ij}\Theta(x_i-x_j)}\right)\nonumber\\
&&\times\frac{1}{\pi^{N-1}}\sum\int_{-\infty}^{+\infty}dx_2\int_{x_2}^{+\infty}dx_3\cdots\int_{x_{N-1}}^{+\infty}dx_N\prod_{i=2}^{N}(x_i^2+1)^{-1}\nonumber\\
&=&\prod_{i<j}e^{-i\pi l_{ij}\Theta(x_i-x_j)}\nonumber\\
&=&f_1*f_2*\cdots *f_N
\labell{star}
\eeqa
where in the first line above we have used the fact 
that the factor
\[
\prod_{i<j}e^{-i\pi l_{ij}\Theta(x_i-x_j)}
\]
depends only on the cyclic ordering of $1,2,3,\cdots,N$\cite{sw}. 
This shows that if one ignores total derivative terms, \ie $p^a$, 
the $*_N$ product reduces to the $*$ product between $N$ open string 
fields. While it was simple exercise  using eq.~\reef{starn},  to show 
the difference between $*_N$ and $*$ product is some total derivative 
terms in the world volume directions, it is   nontrivial calculation to 
show this fact using the integrated form of the $*_N$ as in \reef{star2}.

Using the fact that $\cG^{ij}=N_{ij}$ and $\cG_A^{ab}=-V_A^{ab}$ 
(see Appendix of \cite{mine}), it is not difficult to check that the 
first terms in \reef{hatl1}  
are exactly reproduced by the couplings in \reef{ampthree}.

\section{Non-commutative DBI action}

The string theory calculations \reef{ampthree} shows that the coupling 
between the closed string tachyon and $N$ massless open string 
states when $V_A>>V_S$ is consistent with the following Lagrangian in 
field theory:
\beqa
{\hat{\cL}}(T)&=&-\frac{T_p c}{4}T(\hP,\hA)*_N
\labell{LT}
\eeqa
where
\beqa
T(\hP,\hA)&=&\sum_{n=0,m=0}^{\infty}\frac{\l^{n+m}}{n!m!}\left(\hP^{i_1}\cdots\hP^{i_n}\hA^{a_1}\cdots\hA^{a_m}\right)\nonumber\\
&&(\prt_{x^{i_1}}\cdots\prt_{x^{i_n}})(\prt_{x^{a_1}}\cdots\prt_{x^{a_m}})T^0(x^a,x^i)|_{x^i=0,x^a=\sigma^a}
\labell{ntalor}
\eeqa
and $\hA^a=V_A^{ab}\hA_b$, and the $*_N$ operates as the multiplication 
rules between the $N$ open string fields. If we were able to  evaluate 
the string S-matrix element \reef{amp123} for arbitrary background B-flux, 
then we could  find the couplings between the tachyon and the $N$ open 
string states that involve both $V_A$ and $V_S$. The result should 
reproduce  with  couplings like those in \reef{hatl1}. The 
Lagrangian \reef{LT} and the couplings in \reef{hatl1} are consistent 
with  the following action:
\beqa
\hS&=&-\frac{T_p c}{\sqrt{-\det(V_S)}}\int d^{p+1}\sigma\left(g[T(\hP,\hA)]\sqrt{-\det((V_S)_{ab}+\l \hF_{ab})}\right)*_N
\labell{s1}\eeqa
where the $*_N$ is the multiplication rule between $\hF_{ab}$ and 
the open string fields stemming from  the Taylor expansion \reef{ntalor}. 
This action is of course consistent with the non-commutative DBI action 
\cite{sw} in which there is no closed string field. In that case, 
the $*_N$ operates as the multiplication rule between all the fields 
in the action, hence, using \reef{star} it can be replaced by the $*$ 
product upon ignoring some total derivative terms.

The action \reef{s1}  can still be extended further to include the massless 
closed string fields as well. If one extracts the coupling between one 
graviton and one or two gauge fields from the ordinary DBI action 
\reef{biact}, and then uses the SW map \reef{trans3} to transform them 
to non-commutative counterparts, one will find the following terms:
\beqa
{\hat{\cL}}(h^0,\hA)&=&-\frac{1}{2}T_p c\left( \hA^a\Tr(V\prt_a h^0)-\Tr(Vh^0V\hF)\right)\labell{l12a}\\
{\hat{\cL}}(h^0,2\hA)&=&-\frac{1}{2}T_p c\left(\frac{1}{2}\hA^a*'\hA^b\Tr(V\prt_a\prt_b h^0)-\frac{1}{2}\Tr(Vh^0)\Tr(V_S\hF*'V_S\hF)\right.\nonumber\\
&&\left.\qquad\qquad\qquad-\hA^a*'\Tr(V\prt_a h^0V\hF)+\Tr(Vh^0V\hF*'V_S\hF)\right)
\nonumber
\eeqa
Note that the $V_S^{ab}$, as in \reef{hatl1},  appears only when both 
indices of $V^{ab}$  contract with the open string fields. 
This is consistent with the fact that the $V_S$ is the open string 
metric\cite{sw}. The above terms are consistent with the following 
prescribed action
\beqa
\hS&=&-T_p \int d^{p+1}\sigma\labell{finals}\\
&&\left(g[T(\hP,\hA)]e^{-\phi(\hP,\hA)}\sqrt{-\det\left(P_{\theta}[G_{ab}(\hP,\hA)+B_{ab}(\hP,\hA)]+\l\hF_{ab}\right)}\right)*_N
\nonumber
\eeqa
where now the definition of the pull-back $P_{\theta}$ is the extension 
of \reef{pull} in which ordinary derivative is replaced by its 
non-commutative covariant derivative, \ie, 
$\prt_a\Phi^i\longleftarrow D_a\hP^i=\prt_a\hP^i-i[\hA_a,\hP^i]_M$. 
Expanding above  action around the background B-flux, one finds  
various couplings between open and closed string fields. When both 
indices of $V^{ab}$ contract with the open string fields  
one must replace it with $V_S^{ab}$.
The resulting terms are then fully consistent with the terms 
in \reef{l12a}. When there is no background B-flux, the above 
non-commutative action reduces to the ordinary DBI action \reef{biact}.  
It is also consistent with the non-commutative DBI action in which no 
closed string field is included \cite{sw}. Hence it seems reasonable to 
believe that the above action is the transformation of the ordinary DBI 
action \reef{biact} under the SW map.

The appearance of the $*_N$ and the dependence of the closed string 
fields on the non-commutative gauge fields seems to indicate that the 
non-commutative theory \reef{finals} is not gauge invariant. However, 
the  S-matrix elements calculated in this theory satisfy the Ward 
identity\footnote{The scattering amplitudes are zero under replacing 
polarization of the external states with their momenta.}. Unlike the 
commutative theory \reef{biact} that  contact terms and poles of the 
scattering amplitudes are separately satisfy the Ward identity, in the 
non-commutative case \reef{finals}, however the combination of contact 
terms and the poles of scattering amplitude satisfy the Ward 
identity\cite{mine}.

{\bf Acknowledgments}

I would like to acknowledge 
useful conversation with A.A. Tseytlin and R.C. Myers. 
This work was supported by University of Birjand and IPM.

\newpage
\appendix
\section{The $*_3$ evaluation} 

In this appendix, 
we evaluate the double 
integral that appears in the $*_3$ product \reef{starn}, that is
\beqa
(f_1f_2f_3)*_3&=&\frac{1}{\pi^2}\sum\,e^{-i\pi(l_{12}+l_{13}-l_{23})}\int_{-\infty}^{+\infty}dx_2\int_{x_2}^{+\infty}dx_3\nonumber\\
&&\times (x_2-i)^{-1+l_{21}+l_{23}}(x_2+i)^{-1-l_{21}-l_{23}}(x_3-i)^{-1+l_{31}+l_{32}}(x_3+i)^{-1-l_{31}-l_{32}}\nonumber
\eeqa
In the Appendix of \cite{mrgm}, the following  basic integral is evaluated:
\beqa
L&\equiv &(2i)^{(-3-a-b-c-d-e)}\int_{-\infty}^{+\infty}dx_2\int_{x_2}^{+\infty}dx_3\,
(x_2-i)^a(x_2+i)^b(x_3-i)^c(x_3+i)^d(x_3-x_2)^e
\nonumber\\
&=&-\Gamma(-2-a-b-c-d-e)
\left\{\vphantom{\Gamma\over\Gamma}
(-i)^{2(a+c)}\sin[\pi(b+d+e)]\right.\nonumber\\
&&\qquad\times \frac{\Gamma(-1-d-e)\Gamma(1+e)\Gamma(2+b+d+e)}{\Gamma(-d)
\Gamma(-a-c)}
\nonumber\\
&&\qquad\times {}_3F_2(-c,1+e,2+b+d+e;2+d+e,-a-c;1)\nonumber\\
&&\ +(-i)^{2(a+c+d+e)}\sin(\pi b)\frac{\Gamma(1+d+e)\Gamma(1+b)\Gamma(-1-c-d-e)}
{\Gamma(-c)\Gamma(-1-a-c-d-e)}\nonumber\\
&&\left.\qquad\times {}_3F_2(-d,-1-c-d-e,1+b;-d-e,-1-a-c-d-e;1)
\vphantom{\Gamma\over\Gamma}\right\}\,\,\,.
\nonumber
\eeqa
Using this result, one finds
\beqa
(f_1f_2f_3)*_3&=&-\frac{i}{2\pi^2}\sum\left(\frac{}{}e^{i\pi l_{23}}\sin[\pi(l_{21}+l_{31})]\right.\nonumber\\
&&\times\frac{\Ga(l_{31}+l_{32})\Ga(-l_{21}-l_{31})}{\Ga(1+l_{31}+l_{32})\Ga(2-l_{31}-l_{21})}{}_2F_1(1,-l_{21}-l_{31};2-l_{31}-l_{21};1)\nonumber\\
&&+e^{-i\pi l_{13}}\sin[\pi(l_{21}+l_{23})]\nonumber\\
&&\left.\times\frac{\Ga(-l_{31}-l_{32})\Ga(-l_{21}-l_{23})}{\Ga(1-l_{31}-l_{32})\Ga(2-l_{21}-l_{23})}{}_2F_1(1,-l_{21}-l_{23};2-l_{21}-l_{23};1)\right)
\nonumber
\eeqa
where we have used the fact that 
${}_3F_2(a,b,c;a,d;1)={}_2F_1(b,c;d;1)$. Now using the 
identity ${}_2F_1(1,a;2+a;1)=1+a$, and the properties of the 
Gamma functions, one finds
\beqa
(f_1f_2f_3)*_3&=&\frac{1}{4\pi^2}\sum\left(\frac{e^{i\pi(l_{21}+l_{31}+l_{23})}-e^{-i\pi(l_{21}+l_{31}-l_{23})}}{(l_{31}+l_{32})(l_{31}+l_{21})}+\frac{e^{i\pi(l_{21}+l_{23}-l_{13})}-e^{-i\pi(l_{21}+l_{23}+l_{13})}}{(l_{31}+l_{32})(l_{12}+l_{32})}\right)
\nonumber
\eeqa
after some rearranging the 
right hand side, one finds the result in \reef{star2}.

\end{document}